\documentclass[namedreferences,hyperref,optionalrh]{spr-sola}
\usepackage{graphicx}        
\usepackage{amsmath}        
\usepackage{color}           
\usepackage{lmodern}         

\renewcommand{\vec}[1]{{\mathbfit #1}}

\chardef\us=`\_

\begin{document}

\begin{frontmatter}
\title{Resonant shear-flow instability in anisotropic supersonic plasmas with heat flux}

\author[addressref={aff1,aff2},corref,email={dnamig@gmail.com}]
{\inits{N.S.}\fnm{Namig~S.}~\snm{Dzhalilov}\orcid{0009-0008-3379-8369}}

\address[id=aff1]{Shamakhy Astrophysical Observatory named after N. Tusi, Ministry of Science and Education of the Republic of Azerbaijan, Shamakhy, Azerbaijan}

\address[id=aff2]{Baku State University, Baku, Azerbaijan}

\runningauthor{N. S. Dzhalilov}
\runningtitle{Resonant shear-flow instability in anisotropic supersonic plasmas with heat flux}

\begin{abstract}
This work is devoted to the study of the influence of temperature anisotropy and parallel heat flux on the stability of supersonic shear flow in collisionless plasmas. Within a fluid-based framework, we employ the 16-moment transport equations—derived from the Vlasov–Maxwell system—to describe the plasma dynamics. By performing a modal analysis we investigate the oblique propagation of linear disturbances within a magnetized plasma characterized by a shear flow of arbitrary profile aligned with the ambient magnetic field. In the unperturbed state, both the plasma density and the magnetic field are assumed to be homogeneous. For a smooth, hyperbolic velocity profile representing supersonic shear, the governing wave equation is reduced to a form amenable to an exact analytical solution. Analytical solutions are expressed in terms of special functions that yield an infinite discrete spectrum of complex eigenfrequencies ( $n = 0, 1, 2, \dots $). The instability is identified as resonant, peaking when the wave phase velocity matches the mean flow velocity, with the growth rate decreasing for higher-order modes. The results indicate that, while heat flux exerts a negligible influence under conditions of supersonic flow, the growth rate decreases and approaches an asymptotic value as the Mach number increases. Notably, the instability vanishes in the vortex sheet limit, distinguishing it from the classical Kelvin-Helmholtz mechanism. These findings suggest that this specific instability holds significant potential for explaining the problem of observed boundaries between isotropic and anisotropic proton temperature regions in a low-beta solar wind plasma. 
\end{abstract}
\end{frontmatter}

\section{Introduction}

Shear flows are ubiquitous in astrophysical (e.g., \citealp{Balbus1991, Kiuchi2015,Rieger2019}), geophysical (e.g., \citealp{Hasegawa1975}; \citealp{Farrugia1994}), and laboratory plasma environments (e.g., \citealp{Terry2000}). Defined by a transverse spatial gradient in the velocity field, these flows serve as a fundamental source of multiscale turbulence, which governs mixing, momentum transport, heating, and structure formation (e.g.,\citealp{Maiorano2020,Hillier2024}). The onset of such turbulence is primarily attributed to various shear-driven instabilities (\citealp{Balbus1991}). Consequently, characterizing the transition to turbulence under various physical constraints remains a central problem in plasma dynamics, having received extensive treatment in the literature (e.g, \citealp{Chen1974,Scarf1981,Miura1992,Keller1999,Barranco2009}).

The vortex sheet approximation represents the most elementary framework for modeling shear-driven instabilities (\citealp{Chandrasekhar1961,Gerwin1968}), wherein the velocity gradient is idealized as a delta function. This approach treats the interaction of two distinct media moving with relative constant velocities across subsonic, transonic, supersonic, and relativistic (e.g.,\citealp{Fejer1964,Blandford1976,Ferrari1980,Baty2006}). In this limit, the transition layer is assumed to have a vanishing thickness, forming a stream interface between these two environments where the Kelvin--Helmholtz instability (KHI) develops (e.g.,\citealp{Chandrasekhar1961,Drazin1981}). The characteristics of KHI have been extensively analyzed under various hydrodynamic and magnetohydrodynamic (MHD) conditions (e.g., \citealp{Miura1992}). In the absence of a magnetic field, the KHI is excited for any degree of velocity shear. However, a fluid-aligned magnetic field suppresses the growth rate via magnetic tension. At sufficiently high Mach numbers, the vortex sheet reaches a critical threshold and becomes marginally stable (\citealp{Peyrichon2025}). In partially ionized plasmas, the dynamics exhibits increased complexity (e.g., \citealp{Soler2012}). A fundamental limitation of the vortex sheet model is that the growth rate scales linearly with the wavenumber, implying a physical divergence at short wavelengths (\citealp{Chandrasekhar1961}). As spatial scales approach or fall below characteristic plasma scales, the fluid description breaks down, necessitating a transition to kinetic modeling.

A more sophisticated approach to modeling instability phenomena is the shearing layer approximation, which accounts for a transitional region with finite thickness and a smooth velocity profile between interacting flows (e.g.,  \citealp{Blumen1975, Miura1982, RoyChoudhury1986,Glatzel1988}). The instability dynamics of finite-width shear layers diverge significantly from the idealized vortex sheet limit. This increased complexity arises from the destabilization of the fluid's eigenoscillations by shear flow, triggering phenomena such as mode coupling, interactions with Kelvin--Helmholtz (KH) modes, and various resonant or transient interactions (\citealp{Blumen1975,Ray1982}). In MHD systems, shear interfaces remain susceptible to KH-type instabilities. Analysis by \citet{Belyaev2012} demonstrates that non-magnetic supersonic shear layers exhibit two distinct instability branches: a vortex-sheet-like mode and a sonic-type instability intrinsic to the flow. In magnetized configurations, a fluid-aligned magnetic field also acts as a stabilizing force, suppressing vortex formation via magnetic tension (e.g., \citealp{Faganello2017}). However, if the velocity jump across the interface exceeds the local Alfv\'en speed, this stabilization is overcome, allowing vortex roll-up and the subsequent nonlinear evolution of the instability (e.g., \citealp{Ruffolo2020}).

Linear stability analysis of shear flows typically proceeds via two primary methodologies: modal and non-modal analysis. The classical modal approach employs a Fourier-in-time formalism to determine the eigenfrequency spectrum through the solution of a corresponding boundary value problem (\citealp{Rayleigh1880}). This method characterizes the asymptotic behavior of perturbations over long timescales. However, in shear flow systems governed by non-normal operators, individual eigenmodes may exhibit significant transient growth and interference, even if they are asymptotically stable (e.g.,\citealp{Kaghashvili2000,Rogava2004,Li2006,Shergelashvili2006}). To capture these short-term dynamics, which are neglected by traditional eigenmode analysis, the non-modal approach is employed (e.g, \citealp{Schmid2007,Camporeale2012}). This framework emphasizes the transient evolution of a perturbed equilibrium by treating the wavenumber as a time-dependent variable rather than relying on fixed eigenfrequencies. Extensive research indicates that such transient fluctuations can trigger turbulence and transport even in modally stable systems (\citealp{Schmid2007}). While modal techniques address the long-term stability and asymptotic limits, non-modal methods describe the immediate linear response to perturbations. As noted by \citet{Camporeale2009}, non-normality effects, including mode coupling, become increasingly prominent at kinetic scales and higher plasma beta values. In this study, we adopt the classical modal framework to investigate large-scale phenomena within a fluid description for arbitrary plasma beta values.

Weakly collisional or collisionless, predominantly supersonic plasma flows in magnetized environments are ubiquitous in space and geophysical contexts, including the solar wind, the heliosphere, and the planetary magnetospheres (e.g., \citealp{Borovsky2010,Johnson2014}). These plasmas typically exhibit pronounced temperature anisotropy relative to the ambient magnetic field orientation. Observations consistently reveal marked anisotropy in solar wind populations (e.g.,\citealp{Hellinger2006,Stverak2008,Bale2009,Maruca2011,Osman2012}), where supersonic velocity shears at the interfaces of fast and slow streams serve as primary sites for intense turbulence (\citealp{Verscharen2019}). The KHI in anisotropic regimes has been extensively modeled using the CGL fluid equations (\citealp{Chew1956}) for scenarios where collisions are insufficient to maintain a scalar pressure, but remain frequent enough to suppress heat flow (e.g, \citealp{Talwar1965,RoyChoudhury1986a,Hunana2019}). These studies have characterized the KHI in finite-width shear layers and explored non-local interactions with firehose and mirror instabilities. While KHI growth rates are generally enhanced by anisotropy compared to isotropic MHD, the CGL approximation is fundamentally limited by its reliance on two polytropic closure laws that neglect the anisotropic heat flux inherent to strongly magnetized plasmas. This limitation is addressed by the 16-moment transport equations, derived from the Vlasov kinetic equation (\citealp{Ramos2003,Chus2006}). By incorporating parallel heat flux, this framework provides a more rigorous description of plasma dynamics, enabling a precise analysis of KHI and associated transport phenomena in anisotropic shear flows (e.g.,\citealp{Ismayilli2016b,Ismayilli2016a,Ismayilli2018,Uchava2020}).

This paper investigates the development of shear-driven instabilities in an anisotropic plasma within the framework of the 16-moment fluid equations, accounting for the influence of parallel heat flux. In Section 2, we derive the general governing equations describing the linear phase of the instability for a steady equilibrium flow with an arbitrary velocity profile. In Section 3, we address the case of oblique wave propagation within a transition zone characterized by a smooth hyperbolic velocity profile; here, for the case of supersonic flow, the resulting boundary value problem is solved analytically to determine the complex eigenvalue spectrum. In Section 4, we calculate the instability growth rates using representative solar wind anisotropy parameters as a function of the wavenumber and transition layer width. Finally, Section 5 provides a summary of our findings and a discussion of their physical implications.

\section{The Basic Wave Equations in the Fluid Description}

For the fluid description of a collisionless anisotropic plasma with respect to the direction of the external magnetic field, the 16-moment set of equations may be used, which is complete (in comparison to the CGL approximation) in the sense that these equations include the evolution of heat fluxes along the magnetic field (\citealp{Oraevski1985, Ramos2003}). For the one component (ion) plasma, these equations are given as follows:

\begin{equation}
\frac{d \rho}{dt}+\rho\,\mathrm{div}\,\mathbfit{v}=0,
\label{rho}
\end{equation}
\begin{eqnarray}
\rho \frac{d\mathbfit{v}}{dt}
&+&
\nabla \left( p_{\perp}+\frac{B^{2}}{8\pi} \right) -\frac{1}{4\pi} \left(\mathbfit{B}\cdot\nabla\right)\mathbfit{B} \nonumber\\
&=&
\rho \mathbfit{g} +\left(p_{\perp}-p_{\parallel}\right) \left[\mathbfit{h}_{B}\,\mathrm{div}\,\mathbfit{h}_{B} +\left(\mathbfit{h}_{B}\cdot\nabla\right)\mathbfit{h}_{B} \right] \nonumber\\
&+&
\mathbfit{h}_{B}\left(\mathbfit{h}_{B}\cdot\nabla\right)\left(p_{\perp}-p_{\parallel} \right),
\label{momentum}
\end{eqnarray}
\begin{equation}
\frac{d}{dt}\frac{p_{\parallel}B^{2}}{\rho^{3}}=-\frac{B^{2}}{\rho^{3}}\left[B\left(
\mathbfit{h}_{B}\cdot\nabla \right) \left( \frac{S_{\parallel}}{B} \right) + \frac{2S_{\perp}}{B}
\left( \mathbfit{h}_{B}\cdot\nabla \right)B\right],
\label{ppar}
\end{equation}
\begin{equation}
\frac{d}{dt}\frac{p_{\perp}}{\rho B}=-\frac{B}{\rho}\left(\mathbf{h}_{B}\cdot\nabla\right)
\left(\frac{S_{\perp}}{B^{2}}\right),
\label{pper}
\end{equation}
\begin{equation}
\frac{d}{dt}\frac{S_{\parallel}B^{3}}{\rho^{4}}=-\frac{3p_{\parallel}B^{3}}{\rho^{4}}\left(
\mathbf{h}_{B}\cdot\nabla\right)\left(\frac{p_{\parallel}}{\rho}\right),
\label{Spar}
\end{equation}
\begin{equation}
\frac{d}{dt}\frac{S_{\perp}}{\rho^{2}}=-\frac{p_{\parallel}}{\rho^{2}}\left[\left(
\mathbf{h}_{B}\cdot\nabla\right)\left(\frac{p_{\perp}}{\rho}\right)+\frac{p_{\perp}}{\rho}
\frac{p_{\perp}-p_{\parallel}}{p_{\parallel}B}\left(\mathbf{h}_{B}\cdot\nabla\right)B\right],
\label{Sper}
\end{equation}
\begin{equation}
\frac{d\mathbf{B}}{dt}+\mathbf{B}\,\mathrm{div}\,\mathbfit{v}-\left(\mathbf{B}\cdot\nabla\right)
\mathbfit{v}=0,
\label{B}
\end{equation}
\begin{equation}
\mathrm{div}\,\mathbf{B}= 0,
\label{divB}
\end{equation}
where $\rho$ denotes the density, $p_\parallel$ and $p_\perp$ the parallel and perpendicular gas pressure,
$\mathbf{B}$ the magnetic field, $\mathbfit{v}$ the bulk velocity of the plasma, $\mathbf{g}$ the gravitational acceleration, $\mathbf{h}_{B} = \mathbf{B}/B$ is a unit vector of the magnetic field, and $\frac{d}{dt}=\frac{\partial}{\partial t}+\left(\mathbfit{v}\cdot\nabla\right)$ denotes the convective derivative. Here, $S_\parallel$ and $S_\perp$ are the heat fluxes along the magnetic field due to parallel and perpendicular thermal kinetic motions of ions, respectively. If the heat fluxes are neglected, i.e., when $S_\parallel=0$ and $S_\perp=0$, we obtain with Eqs.~(\ref{rho}) -(\ref{pper}), (\ref{B}) and (\ref{divB}) a closed system of equations that is called
the CGL (Chew-Goldberger-Low) equations, see the pioneering work by \citet{Chew1956}. The closed system of equations (\ref{rho})-(\ref{divB}), which includes the evolution of heat fluxes, is more general than the classical CGL equations.

In our calculations, Eqs.~(\ref{rho})-(\ref{Sper}) are simplified versions of 16-moment equations, including both the ionic and electronic components of the plasma (\citealp{Ramos2003}). Since the plasma is collisionless, electron and ion fluids are weakly coupled. In the condition of $m_e/m_i \ll 1$, the fluid behavior of the plasma is essentially determined by the ionic component. Equations (\ref{rho})-(\ref{divB}), as in the case of the CGL MHD model, describe the dynamics of the ion plasma, while the role of electrons is reduced only to the implementation of the plasma quasi-neutrality condition.

We consider a plane-parallel geometry of the $\mathbf{z}$-directed plasma flow with shearing in the $\mathbf{x}$-direction, i.e. the equilibrium bulk velocity $\mathbfit{v}_{0}$ has a component only in the $\mathbf{z}$-direction, which varies along the $\mathbf{x}$-axis, where $\mathbfit{v}_{0}=\left(0,\,0,\,V_{0}(x)\right)$.
Let the background magnetic field $\mathbfit{B}_{0}$ be directed along the $\mathbf{z}$-axis as well. We also assume that the background state with non-zero heat fluxes is homogeneous, i.e. the gravitational acceleration $\mathbfit{g}=0$, and
$ \left(\rho_{0}, p_{\perp 0}, p_{\parallel 0}, B_{0}, S_{\perp 0}, S_{\parallel 0} \right) = \mathrm{const}$.
To study the stability of such a system to small perturbations, we represent all physical quantities of the basic state in the form
$$
f = f_{0} + f^{\prime}(x)\exp\left[i\left(k_{y}y + k_{z}z - \omega t\right)\right].
$$
For example, the velocity is written as
$\mathbfit{v}=\mathbfit{v}_{0}+\mathbfit{v}^{\prime}(x)\exp\left[i\left(k_{y}y+k_{z}z-
\omega t\right)\right]$, where 
$\mathbfit{v}^{\prime} = \left(\mathrm{v}_{x},\, \mathrm{v}_{y},\, \mathrm{v}_{z}\right)$.
So we Fourier  decompose the perturbation with respect to $y, z$ and $t$, as all the coefficients in the governing  differential equations are dependent on $x$ only. Here, $\omega$ is the wave frequency,  $k_y$ and $k_z$ are the wave numbers, which means that the wave vector $\vec{ k}$ lies in the ($y, z$) plane. After linearization of the set of Eqs.~ (\ref{rho})-(\ref{divB}), we obtain the following

\begin{equation}
\frac{\rho^{\prime}}{\rho_0}=\frac{k_y}{\omega_z} \mathrm{v}_y+\frac{k_z}{\omega_z} \mathrm{v}_z-\frac{i}{\omega_z} \frac{\partial \mathrm{v}_x}{\partial x},
\label{rho'}
\end{equation}

\begin{equation}
\begin{gathered}
\rho_0 \omega_z \mathrm{v}_x+i \frac{\partial p_{\perp}^{\prime}}{\partial x}+i \frac{B_0}{4 \pi} \frac{\partial B_z}{\partial x}+\frac{B_0}{4 \pi} k_z B_x-p_{\Delta} k_z \frac{B_x}{B_0}=0, \\
\end{gathered}
\label{vx}
\end{equation}
\begin{equation}
\rho_0 \omega_z \mathrm{v}_y-k_y\left(p_{\perp}^{\prime}+\frac{B_0}{4 \pi} B_z\right)+\frac{B_0}{4 \pi} k_z B_y-p_{\Delta} k_z \frac{B_y}{B_0}=0 \\,
\label{vy}
\end{equation}
\begin{equation}
\rho_0 \omega_z \mathrm{v}_z+i \rho_0 V_0^{\prime}(x) \mathrm{v}_x+i \frac{p_{\Delta}}{B_0} \frac{\partial B_x}{\partial x}-\frac{p_\Delta}{B_0} k_y B_y-k_z p_{\parallel}^{\prime}=0 \\,
\label{vz}
\end{equation}
\begin{equation}
p_{\perp}^{\prime}=\frac{k_z}{\omega_z}\left(S_{\perp}^{\prime}-2 S_{\perp_0} \frac{B^{\prime}}{B_0}\right)+p_{\perp_0}\left(\frac{B^{\prime}}{B_0}-\frac{\rho^{\prime}}{\rho_0}\right),
\label{pper'}
\end{equation}
\begin{equation}
\begin{aligned}
p_{\parallel}^{\prime}=\frac{k_z}{\omega_z}\left(S_{\parallel}^{\prime}-S_{\parallel_0} \frac{B^{\prime}}{B_0}+2 S_{\perp_0} \frac{B^{\prime}}{B_0}\right)-p_{\parallel_0}\left(2 \frac{B^{\prime}}{B_0}+3 \frac{\rho^{\prime}}{\rho_0}\right),\\
\end{aligned}
\label{ppar'}
\end{equation}

\begin{equation}
\begin{aligned}
S_{\perp}^{\prime}=\frac{p_{\parallel 0} p_{\perp 0} k_z}{\rho_0 \omega_z}\left(\frac{p_{\perp}^{\prime}}{p_{\perp 0}}-\frac{\rho^{\prime}}{\rho_0}+\frac{p_{\Delta} B^{\prime}}{p_{\parallel 0} B_0}\right)+2 S_{\perp_0} \frac{\rho^{\prime}}{\rho_0}, \\
\end{aligned}
\label{Sper'}
\end{equation}

\begin{equation}
\begin{aligned}
S_{\parallel}^{\prime}=3 \frac{p_{\parallel_0}^2 k_z}{\rho_0 \omega_z}\left(\frac{p_{\parallel}^{\prime}}{p_{\parallel_0}}-\frac{\rho^{\prime}}{\rho_0}\right)-S_{\parallel_0}\left(3 \frac{B^{\prime}}{B_0}-4 \frac{\rho^{\prime}}{\rho_0}\right),
\end{aligned}
\label{Spar'}
\end{equation}
\begin{equation}
\begin{gathered}
\omega_z B_x+B_0 k_z \mathrm{v}_x=0, 
\end{gathered}
\label{Bx}
\end{equation}
\begin{equation}
\begin{gathered}
\omega_z B_y+B_0 k_z \mathrm{v}_y=0, 
\end{gathered}
\label{By}
\end{equation}
\begin{equation}
\begin{gathered}
\frac{\partial B_x}{\partial x}+i k_y B_y+i k_z B_z=0 .
\end{gathered}
\label{divB'}
\end{equation}
As evident from formulas (\ref{Sper'}) and (\ref{Spar'}), it is crucial to recognize that even in the ground state, where heat flux is absent ($S_{\perp 0}=S_{\parallel 0} = 0$), the disturbed heat fluxes remain non-zero. This implies that transitioning to the results derived from the CGL equations should be generally avoided.
Inserting  (\ref{Sper'}) and (\ref{Spar'}) into (\ref{pper'}) and (\ref{ppar'}), we can write
\begin{equation}
\begin{aligned}
p_{\perp}^{\prime}=p_{\perp_0} \frac{a_1(x)}{a_0(x)} \frac{B_z}{B_0}+p_{\perp_0} \frac{a_2(x)}{a_0(x)} \frac{\rho^{\prime}}{\rho_0}, 
\end{aligned}
\end{equation}
\begin{equation}
\begin{aligned}
p_{\parallel}^{\prime}=p_{\parallel_0} \frac{b_1(x)}{b_0(x)} \frac{B_z}{B_0}+p_{\parallel_0} \frac{b_2(x)}{b_0(x)} \frac{\rho^{\prime}}{\rho_0} .
\end{aligned}
\end{equation}
Here $p_{\Delta}=p_{\parallel 0}-p_{\perp 0}, V_0^{\prime}(x)=\partial \mathrm{V}_0 / \partial x$ and $\omega_z(x)=\omega-k_z V_0(x) =\omega_D$ is the Doppler shifted frequency in the plasma frame. We can reduce the system of Eqs.~(\ref{rho'})-(\ref{divB'}) to one single second-order differential equation by using the obvious relations between the physical variables
\begin{equation}
\mathrm{v}_x=-\frac{\omega_z}{B_0 k_z} B_x,
\end{equation}

\begin{equation}
\begin{gathered}
\mathrm{v}_y=-\frac{\omega_z}{B_0 k_z} \frac{i k_y A}{\beta_A} \frac{\partial B_x}{\partial x}, 
\end{gathered}
\end{equation}
\begin{equation}
\begin{gathered}
\mathrm{v}_z=-\frac{\omega_z}{B_0} \frac{i A \beta_1}{\beta_{\star}} \frac{\partial B_x}{\partial x}, 
\end{gathered}
\end{equation}
\begin{equation}
\begin{gathered}
B_y=\frac{i k_y A}{\beta_A} \frac{\partial B_x}{\partial x}, 
\end{gathered}
\end{equation}
\begin{equation}
\begin{gathered}
B_z=\frac{i k_z A}{\beta_{\star}} \frac{\partial B_x}{\partial x}, 
\end{gathered}
\end{equation}

\begin{equation}
\begin{gathered}
\rho^{\prime}=\frac{\rho_0}{B_0} \frac{i k_z A\left(1-\beta_1\right)}{\beta_{\star}} \frac{\partial B_x}{\partial x}, 
\end{gathered}
\end{equation}
\begin{equation}
\begin{gathered}
p_{\perp}^{\prime}=\frac{a_1+\left(1-\beta_1\right) a_2}{a_0} \frac{p_{\perp_0}}{B_0} \frac{i k_z A}{\beta_{\star}} \frac{\partial B_x}{\partial x}, 
\end{gathered}
\end{equation}
\begin{equation}
\begin{gathered}
p_{\parallel}^{\prime}=\frac{b_1+\left(1-\beta_1\right) b_2}{b_0} \frac{p_{\parallel_0}}{B_0} \frac{i k_z A}{\beta_{\star}} \frac{\partial B_x}{\partial x} .
\end{gathered}
\end{equation}
Substituting all these relationships and thus eliminating all other variables than $B_x$, we obtain the equation of

\begin{equation}
\frac{\partial}{\partial x}\left(A(x) \frac{\partial B_x}{\partial x}\right)-\beta_A(x) B_x=0, 
\label{eqBx}
\end{equation}
where
\begin{equation}
A(x)=\frac{\beta_A \beta_{\star}}{k_y^2 \beta_{\star}+k_z^2 \beta_A},
\end{equation}
and
\begin{eqnarray}
\beta_{A}
&=&
\beta-\bar{\alpha}-\frac{1}{\eta^{2}},\quad\beta_{\ast}=\beta_{0}-\alpha\frac{a_{2}}{a_{0}}\beta_{1},
\nonumber\\
\beta_{0}
&=&
\alpha\frac{a_{1}+a_{2}}{a_{0}}+\beta,\quad \beta_{1} = \frac{b_{0}\bar{\alpha}-b_{1}-b_{2}}
{b_{0}/\eta^{2}-b_{2}},
\label{beta}
\end{eqnarray}

while the coefficients $a_{0,1,2}$ and $b_{0,1,2}$ are determined as

\begin{eqnarray}
a_{0}
&=&
1-\eta^{2},\quad b_{0} = 1-3\eta^{2},
\nonumber\\
a_{1}
&=&
1-2\gamma\eta-\bar{\alpha}\eta^{2}, \quad b_{1} = 2\gamma\eta(\alpha-2)-2,
\nonumber\\
a_{2}
&=&
1+2\gamma\eta-\eta^{2}, \qquad b_{2} = 3+4\gamma\eta-3\eta^{2}\,.
\label{coeff}
\end{eqnarray}
In the above equations, the following dimensionless parameters were used for the basic unperturbed physical quantities, and the zero indices are dropped for simplification of the notation:
\begin{eqnarray}
\alpha
&=&
\frac{p_{\perp}}{p_{\parallel}} = \frac{T_{\perp}}{T_{\parallel}}, \quad \bar{\alpha} =
1-\alpha, \quad \beta = \frac{B^{2}}{4\pi p_{\parallel}} = \frac{v_{A}^{2}}{c_{\parallel}^{2}},
\nonumber\\
c_{\parallel}^{2}
&=&
\frac{p_{\parallel}}{\rho},\quad \eta = \frac{c_{\parallel}k_{z}}{\omega_{z}}, \quad
\gamma_{\parallel} = \frac{S_{\parallel}}{p_{\parallel}c_{\parallel}},
\gamma_{\perp} = \frac{S_{\perp}}{p_{\perp}c_{\parallel}},
\gamma = \gamma_{\parallel} = \gamma_{\perp}\,.
\label{parameters}
\end{eqnarray}
The case where $\gamma=\gamma_{\parallel}=\gamma_{\perp}$ is a simplified approach.
Here, $c_{\parallel}$ denotes the parallel sound speed, $v_A$ is the Alfv\'en velocity,
$\alpha$ is the anisotropy parameter, $\gamma$ the background heat flux parameter, and the magnetic field parameter $\beta$ is inversely proportional to the plasma beta, $\beta=2/\beta_p$. We also denote $\theta$ for the wave propagation angle relative
to the magnetic field, i.e., $k_z=k\cos\theta$, $k_y=k\sin\theta$, and after denoting
$\cos^2\theta=\ell$, we can write $k_z=k\sqrt{\ell}$ and $k_y=k\sqrt{1-\ell}$, and let $k>0$.

\section{Supersonic Shearing Flow Instability}

\begin{figure}
  \centering
\includegraphics[width=0.9\textwidth]{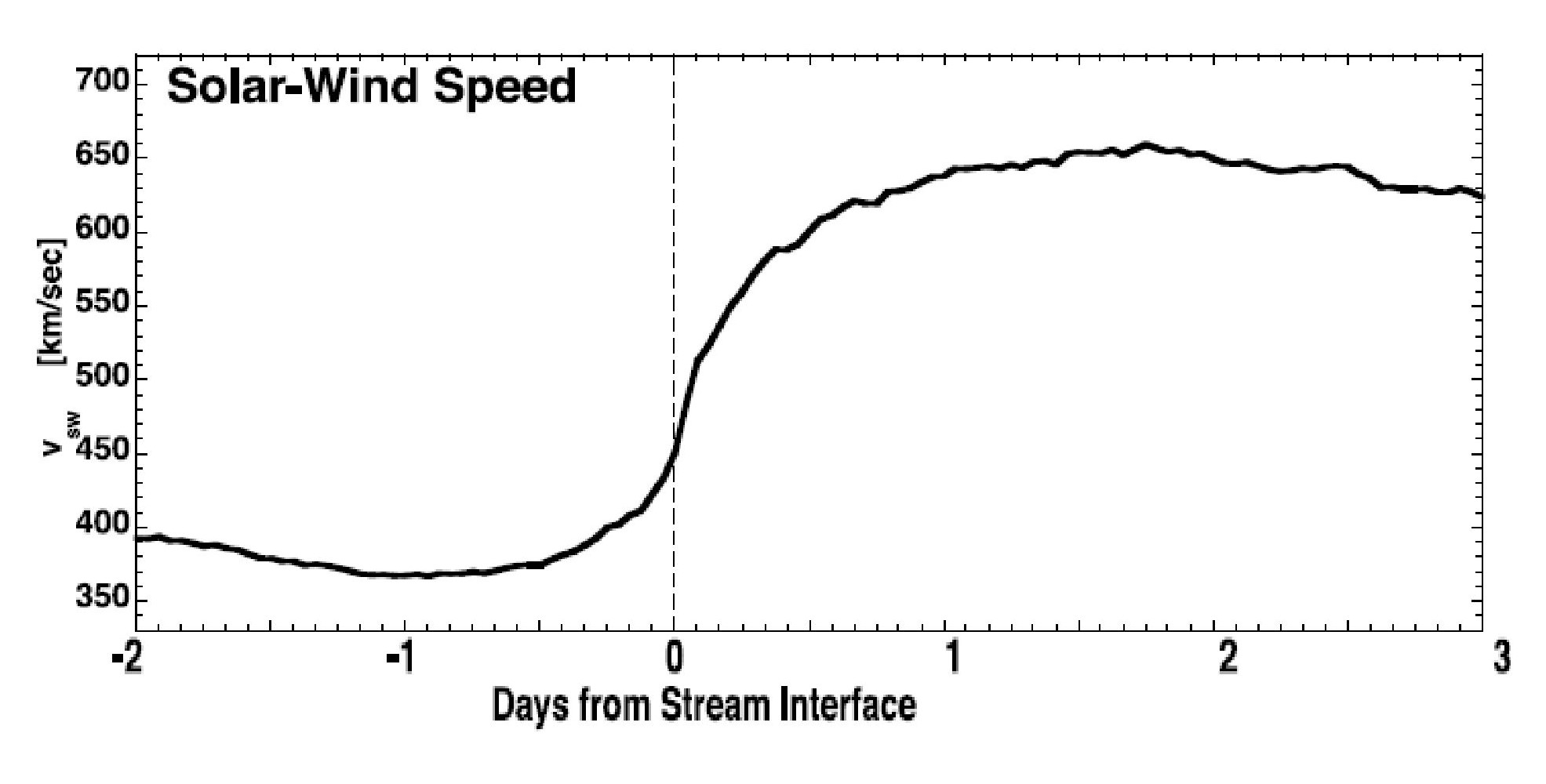}
  \caption{Overview of the solar wind corotating interaction regions (CIR) structure. The superposed averages of OMNI2 measurements for the 27 CIR events flow velocities  of the solar wind are plotted as a function of time. The zero epoch is the CIR stream interface as determined by the maximum in the plasma vorticity. The plots extend from 2 days prior to the passage of the stream interface (in the slow wind) to 3 days afterward (in the fast wind). The rise in the solar wind speed associated with the passage of the CIR. In the slow wind it has been argued that the turbulence is well developed, with fluctuations that propagate both inward and outward. In a CIR the magnetic field lies in the plane of the stream interface. [Reproduced from first panel of Figure 4 in: \citet{Borovsky2010}, Solar wind turbulence and shear: A superposed‐epoch analysis of corotating interaction regions at 1 AU, J. Geophys. Res., 115, A10101, doi:10.1029/2009JA014966. \copyright 2010 American Geophysical Union. Used with permission: license number 6250061247385] }
\label{fig1}
\end{figure}

To address the issue of instability within the system under study, it is necessary to solve the derived wave equation.
The coefficients of this second order ordinary differential equation (\ref{eqBx}) are variable, and they are complex functions of $V_0(x)$. Furthermore, the velocity profile $V_0(x)$ is still an arbitrary function of the $x$-coordinate. As usually, when two plasma flows adjoin, there arises a smooth transition layer with finite thickness between them. A typical example of such a situation is the formation of a contact layer between the slow and fast components of the solar wind. Figure \ref{fig1}, which is borrowed from \citet{Borovsky2010}, shows such corotating interaction regions (CIRs) near Earth's orbit based on space measurements. The thickness of the transition layer is determined by the flow velocities and physical parameters of each flow, which will vary depending on the distance from the Sun. To address equation (\ref{eqBx}), it is essential to first define the velocity profile $V_0(x)$. In many cases, this profile is represented as a linear function of $x$. However, in this particular approach, internal reflections of the propagating waves arise at the boundaries of the linear profile, where solutions are required to be matched. This leads to artificial and physically implausible outcomes that do not occur when smooth profiles are used. In the simplest case, to model this situation, we consider an analytical velocity profile given by the smooth hyperbolic function of 
\begin{equation}
V_{0}(x) = \frac{ V_{02}e^{\sigma x} + V_{01}e^{-\sigma x} }{e^{\sigma x} + e^{-\sigma x}},
\quad \sigma \ge 0 \,.
\label{V0}
\end{equation}
Here $V_0(-\infty) = V_{01}$ and $V_0(+\infty) = V_{02}$ are the limit velocities and let
$h = V_{01}/V_{02} \ge 1$, $V_0(0) = (V_{01} + V_{02})/2 = \overline{V}_0$ is the average flow velocity. In Eqs.~(\ref{V0}) the $\sigma$ parameter characterizes the inverse value of the width of the transition layer $L$. 
By choosing the appropriate parameter values for $V_{01}, V_{02}$ and $\sigma$ in the profile (\ref{V0}), the CIR profile illustrated in figure~\ref{fig1} can be accurately described.
As the parameter $\sigma L > 0$ increases, the width of the transition layer between the two flows decreases sharply and becomes a discontinuity ($\sigma L \gg 1$, vortex sheet interface approximation) between the two flows with the velocities of $V_{01}$ and $V_{02}$. For small $\sigma L \ll 1$ the width of the transition layer becomes very large and the effects of the vortex sheet interface are expected to vanish.

Let
\begin{eqnarray}
y(x)
&=&
\frac{B_{x}}{B_{0}},\quad \xi(x) = \frac{\omega-k_{z}V_{0}(x)}{c_{\parallel}k_{z}},
\nonumber\\
\beta_{A}(\xi)
&=&
\alpha+\beta-1-\xi^{2},
\nonumber\\
\beta_{\ast}(\xi)
&=&
\beta + 2\alpha + 2\alpha^{2} \frac{ \xi^{4} + 2\gamma\xi^{3} + 2\gamma^{2}\xi^{2} - 5\xi^{2} -
6\gamma\xi + 3 }{\left(\xi^{2}-1\right) \left(\xi^{4}-6\xi^{2}-4\gamma\xi+3\right) }\,.
\label{definitions}
\end{eqnarray}

Then we can write Equation~(\ref{eqBx}) as
\begin{equation}
\frac{d}{dx} \left[\frac{\beta_{A}\beta_{\ast}}{(1-\ell)\beta_{\ast}+\ell\beta_{A}}\frac{dy(x)}{dx}\right]- k^{2}\beta_{A}y(x) = 0\,.
\label{eq2}
\end{equation}
It should be noted that in the work \citet{Ismayilli2018} we have shown in a general form that the instability of the considered oscillations occurs under conditions where the phase velocity of the wave along the flow is close to the mean flow velocity $\mathrm{Re}(\omega) \approx k_z \overline{V}_0$. Based on this, we introduce the complex spectral parameter $\Omega$. Let us
\begin{equation}
\omega = k_z \overline{V}_0(1 + \Omega).
\end{equation}
Then we obtain $\xi(x) = M[\Omega+\Delta\, \mathrm{tanh}(\sigma x)]$. Here, $M={\overline{V}_0}/{c_\parallel}$ is the sonic Mach number and $\Delta=(V_{01}-V_{02})/(V_{01}+V_{02})$ is the shear rate of the flow.

Our main goal in the next step is to solve the boundary value problem based on Equation~(\ref{eq2}) to determine the complex spectral parameter $\Omega$ with $\Omega_{i}=\mathrm{Im}(\Omega)\neq0$. In the general case, it is not possible to solve Equation~(\ref{eq2}) analytically. However, some important particular cases of this equation have been studied in our previous works (\citealp{Ismayilli2018,Dzhalilov2023,Dzhalilov2024}). Here, we consider the case of oblique wave propagation ($\ell \neq 0$ and $\ell \neq 1$) in a supersonic shear flow, $M \gg 1$. Then, for large $\xi$, Equation~(\ref{eq2}) is significantly simplified and reduces to the Gauss hypergeometric equation:
\begin{equation}
 y^{\prime \prime }(x)+q^2\left[\Omega+\Delta \tanh \left(\sigma x\right)\right]^2 y(x)=0,
 \label{Gauss}
\end{equation}
where $q^2 = M^2k^2\ell/(\beta+2\alpha)$. Two independent solutions of this equation are expressed in terms of the well-known Gauss hypergeometric functions (\citealp{Bateman1953}):
\begin{equation}
y(x)=w(x)\left[C_1 \zeta^{i \varphi_2} F_1(\zeta)+C_2 \zeta^{-i \varphi_2} F_2(\zeta)\right],
 \label{sols}
\end{equation}
where
\begin{eqnarray}
\varphi_{1}
&=&
\frac{\mu}{2}(\Omega+\Delta),\, \varphi_{2} = \frac{\mu}{2}(\Omega-\Delta), \, \mu =
\frac{q}{\sigma} = M\frac{k_{z}}{\sigma} \frac{1}{\sqrt{\beta+2\alpha}}>0,
\nonumber\\
F_{1}(\zeta)
&=&
\,_{2}F_{1}(a_{1},b_{1};c_{1};\zeta),\quad F_{2}(\zeta) = \,_{2}F_{1}(a_{2},b_{2};c_{2};\zeta),
\nonumber\\
2a_{1}
&=&
1-i\left(\sqrt{4\mu^{2}\Delta^{2}-1}-2\mu\Omega\right),\, 2b_{1} =
1+i\left(\sqrt{4\mu^{2}\Delta^{2}-1}+2\mu\Omega\right),
\nonumber\\
2a_{2}
&=&
1-i\left(\sqrt{4\mu^{2}\Delta^{2}-1}-2\mu\Delta\right), \, 2b_{2} = 1+i\left(\sqrt{4\mu^{2}\Delta^{2}-1}+2\mu\Delta\right),
\nonumber\\
c_{1}
&=&
1+i\mu(\Omega-\Delta),\, c_{2} = 1+i\mu(\Delta-\Omega)\,.
\label{hypergeom-def}
\end{eqnarray}
The arbitrary integration constants $C_1$ and $C_2$ must be determined from the boundary conditions. The boundary conditions are the boundedness of the solution $y(x)$ in the interval $x \to [-\infty, +\infty]$.

In the limit of $x \to +\infty$ we have
$$
\zeta \rightarrow 1-e^{-2 \sigma x}=1-\varepsilon; \ \ \varepsilon=e^{-2 \sigma|x|} \ll 1; \ \ w\approx \frac{i}{2}\left(-e^{-2 \sigma x}\right)^{i \varphi_{_1}}.
$$
Since the complex spectral parameter $\Omega = \mathrm{Re}(\Omega) + i\,\mathrm{Im}(\Omega) = \Omega_r + i\Omega_i$, and we are interested in instability, it must be $\Omega_i > 0$. Then $\varphi_1=\varphi^{+}+i \varphi_*, \  \varphi_2={\varphi}^{-}+i \varphi_*$ and
$$
\varphi^{ \pm}=\frac{\mu}{2}\left(\Omega_r \pm \Delta\right), \varphi_*=\frac{\mu}{2} \Omega_i.
$$
This means that as $x \to +\infty$ we obtain $w \sim e^{2\sigma\varphi_* x} \to \infty$. Therefore, from the requirement of boundedness of the solution, we find that the following condition must be satisfied:
\begin{equation}
\left(C_1 \zeta^{i \varphi_2} F_1(\zeta)+C_2 \zeta^{-i \varphi_2} F_2(\zeta)\right)_{\zeta \rightarrow 1-\varepsilon}=0.
\label{cond1}
\end{equation}

In the limit of $x \rightarrow-\infty$ we have $\zeta \rightarrow e^{2 \sigma x}=\varepsilon$. Then 
$$w \rightarrow const, \ \ \zeta^{i \varphi_2} \rightarrow e^{-2 \varphi_* \sigma x} \rightarrow \infty ; \quad \zeta^{-i \varphi_2} \rightarrow e^{2 \varphi_* \sigma x} \rightarrow 0. \quad 
$$
From the requirement of the boundedness of the solution at $x \to -\infty$, we obtain $C_{1}=0$. Then condition~(\ref{cond1}) reduces to
\begin{equation}
    F_2(\zeta)\big|_{\zeta = 1 - \varepsilon} = 0.
\label{cond2}
\end{equation}
This is the desired dispersion equation:
\begin{eqnarray}
F(p,\eta)
&=&
\,_{2}F_{1}\left(a,b;c;1-\varepsilon\right)=0,
\label{dis}
\\
a
&=&
\frac{1}{2}\left[1-i\left(\sqrt{p^{2}-1}-p\right)\right],
\nonumber\\
b
&=&
\frac{1}{2} \left[1+i\left( \sqrt{p^{2}-1} + p \right) \right],
\nonumber\\
c
&=&
1+i\left(\frac{p}{2}-\eta\right),\quad p=2\mu\Delta, \quad \eta=\mu\Omega\,.
\nonumber
\end{eqnarray}
Thus, the solution of the formulated problem reduces to finding the eigenvalues $\eta$ as a function of the generalized parameter $p$, $F(p,\eta) = 0$. 
Note that for $\varepsilon = 0$, the dispersion equation is expressed in terms of Gamma functions $\tilde\Gamma(..)$:
\begin{equation}
{ }_2 F_1(a,b;c;1)=\frac{\tilde\Gamma(c)\tilde\Gamma(c-a-b)}{\tilde\Gamma(c-a)\tilde\Gamma(c-b)},
\end{equation}
if conditions $\mathrm{Re}(c)>\mathrm{Re}(b)>0$ and $\mathrm{Re}(c-a-b)>0$ are satisfied (\citealp{Bateman1953}). Since these conditions are additional requirements, we shall investigate the more general dispersion equation~(\ref{dis}) for a very small value of $\varepsilon = 10^{-8}$.

Note that since we consider waves in an unbounded medium $-\infty < x < +\infty$, the aperiodic oscillations ($\mathrm{Re}(\omega) = 0$ or $\mathrm{Re}(\Omega) = -1$) represent a special case for special values of the parameter $p$ (see below). For periodic oscillations $\mathrm{Re}(\omega) \neq 0$, the rate of growth of the instability determines the rate of amplitude growth over one oscillation period:
\begin{equation}
\Gamma = \frac{\mathrm{Im}(\omega)}{\mathrm{Re}(\omega)} =
\frac{\mathrm{Im}(\Omega)} {1+\mathrm{Re}(\Omega)} =
\frac{\mathrm{Im}(\eta)} {\mu+\mathrm{Re}(\eta)} \,.
\label{gamma}
\end{equation}

Without loss of generality, we consider the case $\Delta > 0$, i.e. 
$\frac{dV_{0}(x)}{dx}<0 \,.$

\begin{figure}
  \centering
  \includegraphics[width=0.5\textwidth]{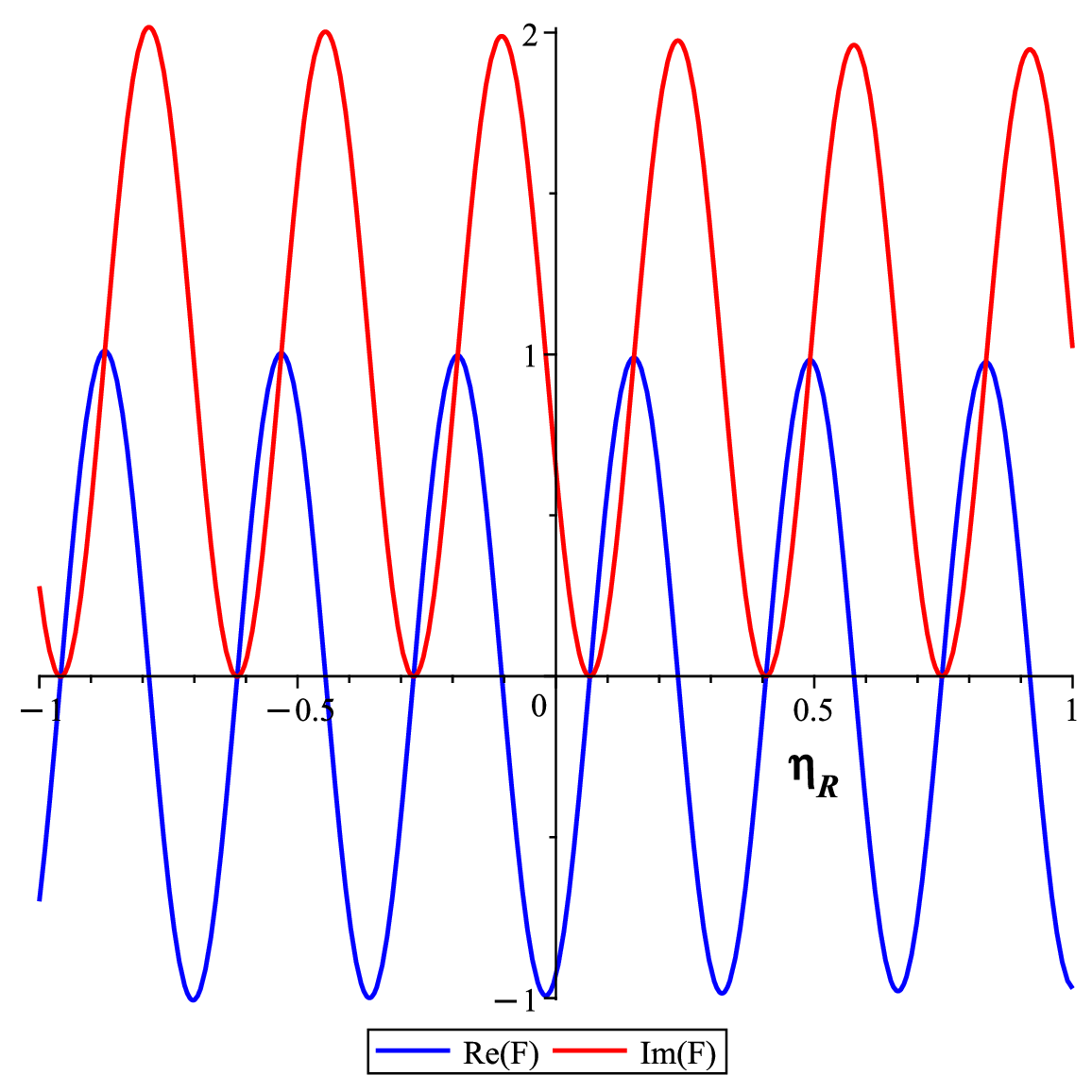}
  \caption{Asymptotic behaviour of the complex dispersion function $F(p,\eta)$ with respect to the real part of the spectral parameter $\eta$ ($\eta_{R}=\mathrm{Re}(\eta)$) for the case of large $p$ (here $p=100$) and $\eta_{i}=\mathrm{Im}(\eta)=0.02$. The simultaneous condition $\mathrm{Re}(F)=0$ and $\mathrm{Im}(F)=0$ corresponds to the existence of an infinite sequence of eigenvalues.}
\label{fig2}
\end{figure}

\begin{figure}
  \centering
  \includegraphics[width=0.5\textwidth]{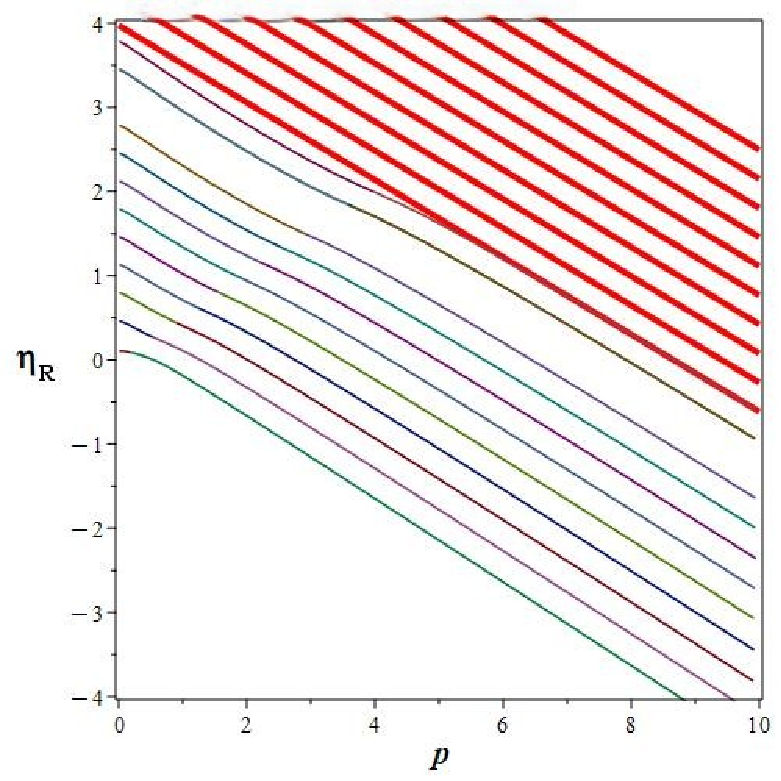}
  \caption{Exact numerical solutions of the dispersion equation $F(p,\eta)=0$ as a function of the parameter $p$. The real parts of the eigenvalues $\eta_R=\mathrm{Re}(\eta)$ are plotted versus $p$. The curves $\eta_R(p)$ correspond to individual modes $n=0,1,2,3,\ldots$ (from bottom to top, respectively). For higher modes ($n \geq 10$), the dependencies $\eta_R(p)$ become linear (red lines) and coincide exactly with the asymptotic analytical expression (\ref{etar}).}
\label{fig3}
\end{figure}

\begin{figure}
  \centering
  \includegraphics[width=0.5\textwidth]{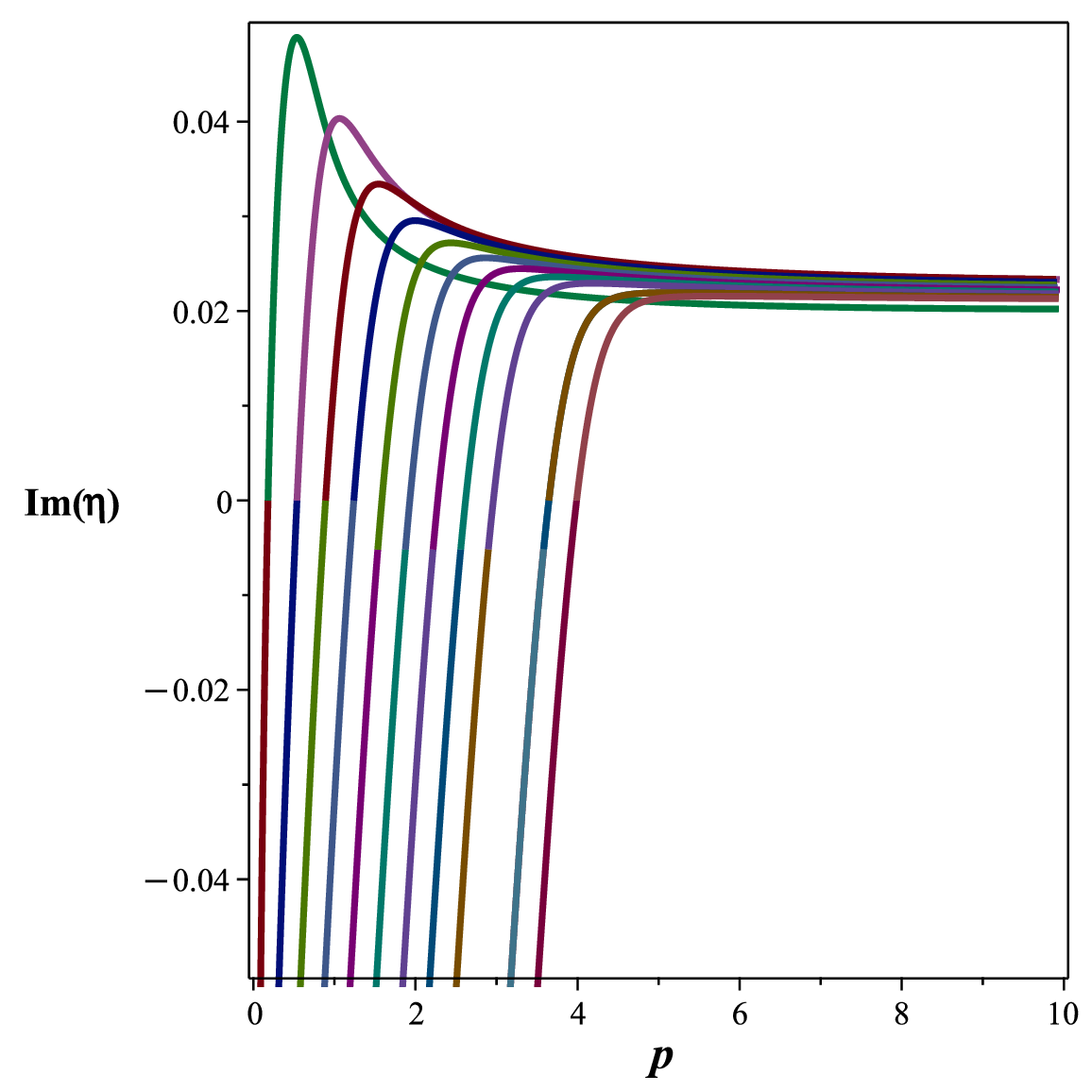}
  \caption{Exact numerical solutions of the dispersion equation $F(p,\eta)=0$ as a function of the parameter $p$. The imaginary parts of the eigenvalues $\eta_i=\mathrm{Im}(\eta)$ are plotted versus $p$. The curves from left to right correspond to the first ten modes $n=0,1,2,\ldots,10$. For higher modes, instability ($\eta_i>0$) appears at larger values of the parameter $p$. For $p>5$, the growth rate rapidly approaches the asymptotic value $\eta_i \rightarrow 0.02$. For all modes there exists a critical value $p_c$ (threshold) such that instability does not occur when $p<p_c$. As the harmonic number increases, the critical value $p_c$ also increases.}
\label{fig4}
\end{figure}

\begin{figure}
  \centering
  \includegraphics[width=0.5\textwidth]{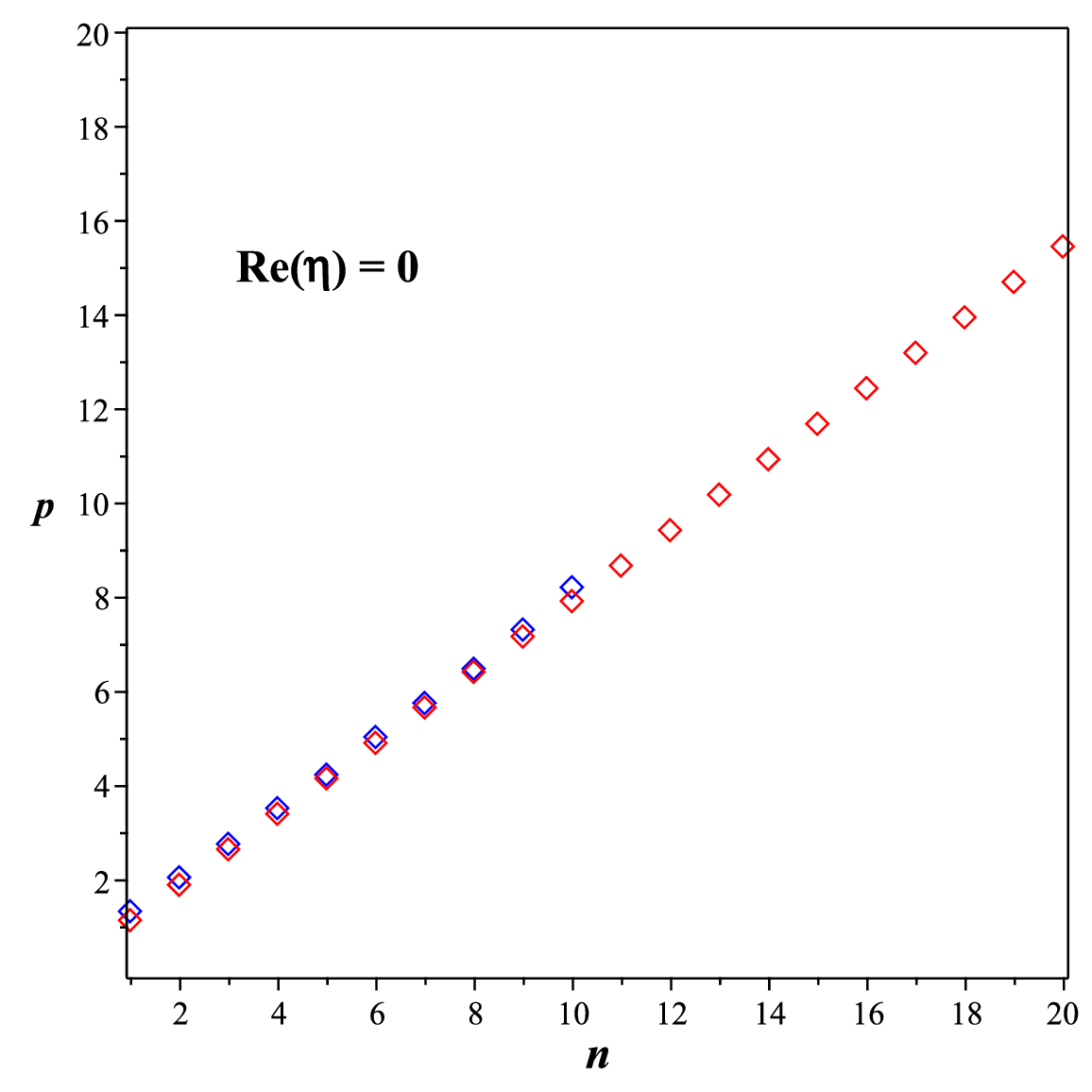}
  \caption{The wave-flow resonance condition, $\eta_{R}(p,n) = 0$ is represented on the plot. The curve illustrates the values of the parameter $ p$ for the first 20 modes, $n = 0, 1, \cdots, 20$. The locations of the points based on the exact solution (\ref{dis}) and the asymptotic formula (\ref{etar}) are practically identical.}
\label{fig5}
\end{figure}

It can be easily demonstrated that the equation $F(p,\eta)=0$ possesses an infinite number of solutions. As the parameter $p$ increases, $\eta_{i}=\mathrm{Im}(\eta)$ tends to its constant limit, $\eta_{i}\approx 0.02$, while $\eta_{R}=\mathrm{Re}(\eta)$ takes infinitely many values.
In Figure~\ref{fig2}, the case $p=100$ is shown for the real and imaginary parts of the dispersion function $F(p,\eta_{R}+0.02i)$. As can be seen, $\mathrm{Re}(F)$ and $\mathrm{Im}(F)$ have an ideal sinusoidal dependence on $\eta_{R}$ with an infinite number of zeros, where $\mathrm{Re}(F)=\mathrm{Im}(F)=0$.
Using similar plots for different large values $p$, we establish the following simple law for the distribution of roots:
\begin{equation}
    \eta_{_R} = 0.5\,p_n - 0.459015\,(p-p_n),
\label{etar}
\end{equation}
where $p_n = p_0(2n + 1)$, $p_0 = 0.1801944$, and the positive integers $n = 0, 1, 2, 3, \ldots$ are the mode numbers.
In figures \ref{fig3} and \ref{fig4}, the exact numerical solutions of the dispersion equation (\ref{dis}) are presented as functions of the parameter $p$. It can be seen that the curves $\eta_R(p)$ are very close to inclined straight lines. In the same figure, straight lines according to formula (\ref{etar}) for $n > 10$ (from bottom to top) are superimposed in red. We observe that starting from $n = 10$, the modes are well described by formula (\ref{etar}). From these figures, the following important facts are obtained:

1) The instability (the region $\eta_i > 0$) has a threshold at small values of $p=p_c$. It is easy to show that for all modes the thresholds correspond to $p_c = p_n$, $\eta_R = p_n/2$. For the first fundamental mode ($n=0$) with the most intensive instability,
\begin{equation}
p_{c}=p_{0}\approx 0.18,
\qquad
\eta_{R}\approx 0.09 \,.
\end{equation}
2) All modes $n$ have zeros, $\eta_R(p)=0$. This corresponds to the resonance of waves with the flow. Since $\omega = k_z \overline{V}_0(1 + \Omega)$, and the resonance with the flow occurs when $\mathrm{Re}(\Omega) \approx 0$ (\citealp{Ismayilli2018}), this corresponds to $\eta_R = 0$. In figure~\ref{fig5} the zero points for the modes $0 \le n \le 20$ are presented based on the exact solutions of the dispersion equation.
The zeros $\eta_R$ according to the formula (\ref{etar}) are also plotted there. We see that these points coincide exactly. Therefore, the resonance condition, as follows from (\ref{etar}), at $\eta_R = 0$ for all $n$ modes, is $p \approx 2.08\, p_n$. For $n = 0$, $p = 0.38$; for $n = 1$, $p = 1.097$; for $n = 2$, $p = 1.188$, etc.

\begin{figure}
  \centering
  \includegraphics[width=0.5\textwidth]{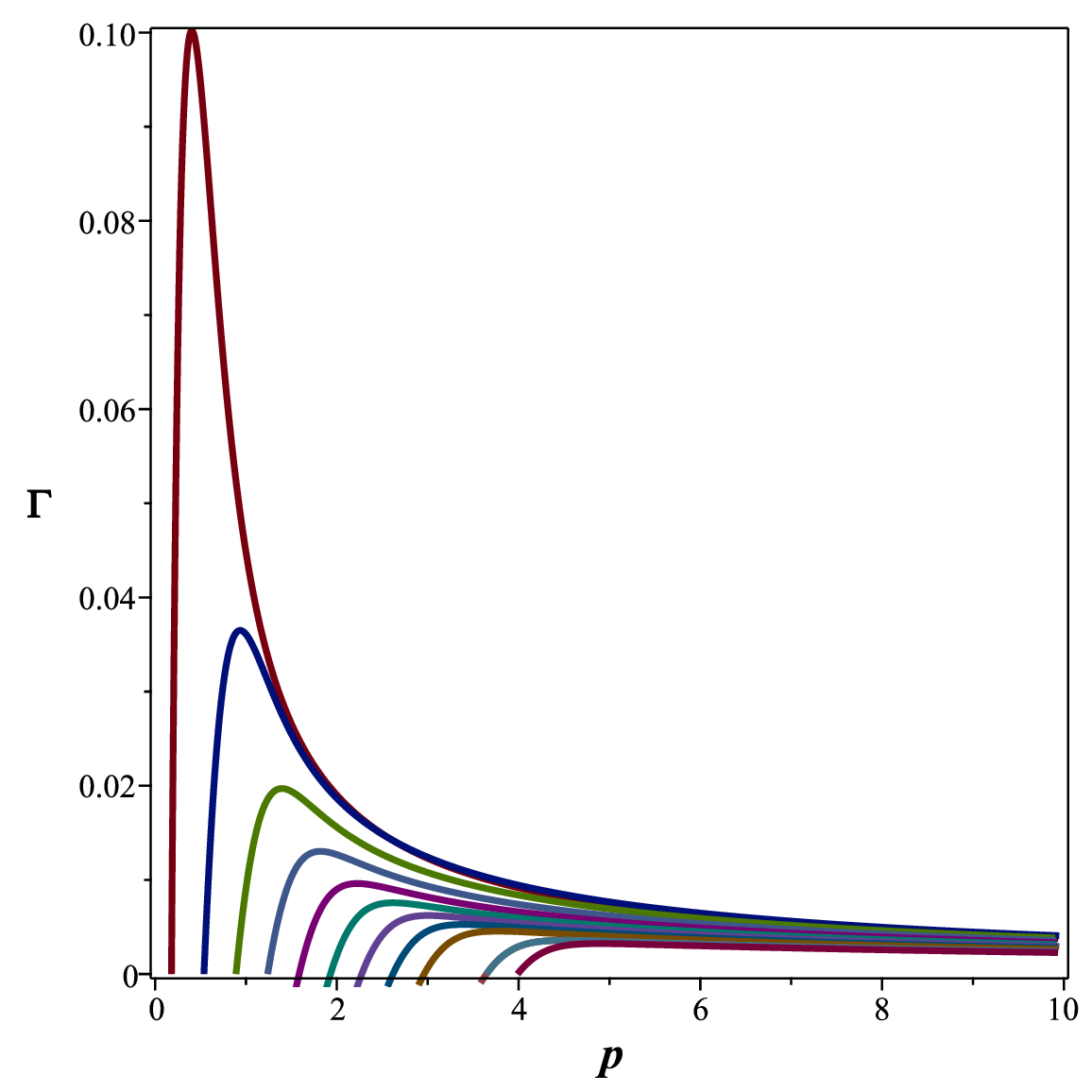}
  \caption{Growth rates $\Gamma(p)$ of the first ten modes ($n = 0,1,\ldots,10$). The maxima correspond to the resonance conditions of the modes with the mean flow.}
\label{fig6}
\end{figure}

3) For each unstable mode, the growth rate has one maximum. In figure~\ref{fig6}, the growth rates $\Gamma(p)$ of the first 10 modes are shown according to formula (\ref{gamma}) ($\mu = p/2\Delta$ and the case of $\Delta = 0.5$). Maksimal growth rates occur under the resonance conditions of the modes with the flow.

4) Aperiodic instability when $\mathrm{Re}(\omega) = 0$, $\mathrm{Im}(\omega) \neq 0$, cannot arise. For such modes to occur, the condition of $\mathrm{Re}(\Omega) = -1$ or $\eta_R = -p/2\Delta$ must be satisfied. Since $0 < \Delta < 1$, this condition is not satisfied according to (\ref{etar}).

\section{On the Problem of Solar Wind Temperature Anisotropy}

\begin{figure}
  \centering
\includegraphics[width=0.5\textwidth]{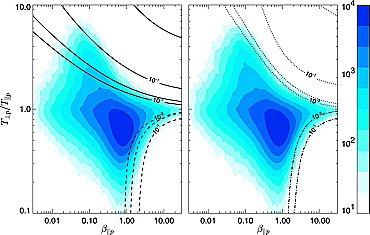}
  \caption{A color scale plot of the relative frequency of  ($\beta_{\parallel p}, T_{\perp p} /T_{\parallel p }$) in the WIND/SWE data (1995–2001) for the solar wind velocity  600 km/s. The (logarithmic) color scale is show on the right. The over plotted curves show the contours of the maximum normed growth rate in the corresponding bi-Maxwellian plasma (left) for the proton cyclotron instability (solid curves) and the parallel fire hose (dashed curves) and (right) for the mirror instability (dotted curves) and the oblique fire hose (dash-dotted curves). [Reproduced from the Figure 1 in: \citet{Hellinger2006}, Solar wind proton temperature anisotropy: Linear theory and WIND/SWE observations. GeoRL, 33, L09101 doi:10.1029/2006GL025925. \copyright 2006 American Geophysical Union. Used with permission: license number 6250070828959].}
\label{fig7}
\end{figure}

Solar wind protons consistently exhibit temperature anisotropies, as highlighted in numerous studies (e.g.,\citealp{Kasper2003,Hellinger2006, Bale2009, Maruca2012, Seough2013,Huang2020}). The adiabatic expansion of the solar wind predicts substantial deviations from temperature isotropy during its outspreading. For instance, an initially isotropic plasma parcel from the outer corona is anticipated to develop a temperature anisotropy of $\alpha = T_\perp/T_\parallel = 0.03$ by the time it reaches 1 AU from the Sun, where $T_\perp$ and $T_\parallel$ denote the perpendicular and parallel temperatures with respect to the ambient magnetic field (\citealp{Phillips1990}). Nonetheless, in situ observations made, for example, by the Wind spacecraft indicate that the actual temperature anisotropy at 1 AU remains confined within certain limits: $0.1 < \alpha < 10$ for protons and $0.5 < \alpha< 2$ for electrons. When represented in histograms that show proton anisotropy $\alpha$ as a function of the parallel plasma beta $\beta_p$ (where $\beta_{\parallel p} = 8\pi n k_B T_ \parallel/B^2 =\beta_p$, with $B$ being the ambient magnetic field strength, $n$ the number of density and $k_B$ the Boltzmann constant), the data form a characteristic rhombic-shaped region, concentrated around $\beta_p = 1$. The dependence of proton temperature anisotropy on $\beta_{\parallel p}$, is presented in figure~\ref{fig7}, which is borrowed from \citet{Hellinger2006}. As can be seen in the figure, the rhomboidal region $(\alpha, \beta_{\parallel p})$ of anisotropy is bounded within $10^{-3} \le \beta_{\parallel p} \le 12,\quad 0.1 < \alpha < 10.$
Several theoretical models aim to explain the observed boundaries $\alpha$ between the isotropic and anisotropic regions. The main idea is that plasma instabilities associated with plasma anisotropy have maximum growth rates at these boundaries, and as a result of the further development of instabilities, the plasma becomes isotropized. In non-streaming bi-Maxwellian plasmas, the linear theory of temperature anisotropy-driven kinetic instabilities, such as mirror, firehose, and electromagnetic cyclotron instabilities, is often applied to interpret these phenomena (e.g.,\citealp{Hellinger2006, Stverak2008, Bale2009, Maruca2011, Lazar2017, Zhao2019, Martinovic2021}). However, these instability thresholds predominantly constrain temperature anisotropy in high $\beta_p > 1$ plasmas, whereas the boundaries in the low-beta regime—particularly those to the left—remain largely unexamined despite various hypotheses (e.g.,\citealp{Matteini2012, Cranmer2014, Ozak2015, Chen2016}). 

Recent studies revisiting anisotropy instabilities in magnetized, (counter-) streaming bi-Maxwellian plasmas highlight that differential streams can markedly modify instability conditions (e.g.,\citealp{Vafin2018}). Significant plasma flow has been established to be likely essential for instabilities at small beta values: the $\alpha > 1$ and $\beta_p < 1$ boundaries arise due to the parallel-propagating left-hand polarized Alfvén wave instability (\citealp{Schlickeiser2010}), whereas the $\alpha < 1$ and $\beta_p < 1$ boundaries are governed by the perpendicular ordinary mode instability (\citealp{Ibscher2014}). Additional insights have emerged from hybrid Vlasov-Maxwell simulations by \citet{Servidio2014}. Beginning with an isotropic turbulent maxwellian plasma ($\alpha = 1$) as their initial state, these simulations demonstrate that plasma turbulence self-consistently generates temperature anisotropy patterns similar to those observed in the solar wind as the plasma evolves over time. Furthermore, collisional relaxation mechanisms are suspected to play a key role in shaping the left-hand boundary $\beta_p < 1$ (e.g.,\citealp{Vafin2019, Yoon2024}). Despite these advances, understanding of the temperature anisotropy in the low $\beta_p$ region remains incomplete and continues to be an active field of research.

Thus, according to most opinions, the boundaries at small beta are most likely associated with flow streaming effects of anisotropic plasma. We will aim to utilize the results obtained here, derived from the fluid description of the anisotropic plasma flow, to explain the formation of these boundaries at $\beta_{p} < 1$.
\begin{figure}
  \centering
  \includegraphics[width=0.5\textwidth]{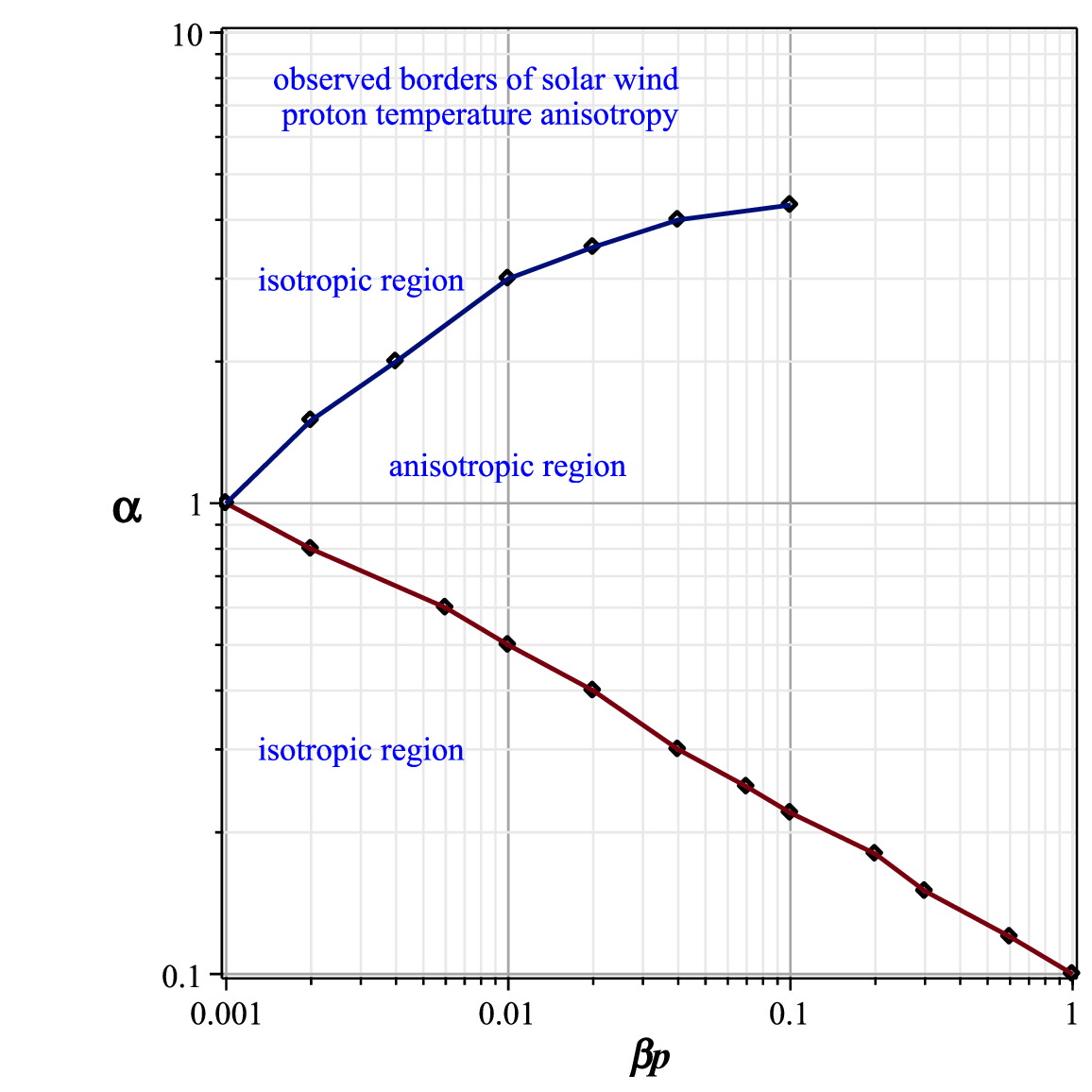}
  \caption{The curves representing the left boundaries, extracted from the figure \ref{fig7}, indicate the transition between isotropic and anisotropic regions for low plasma beta conditions, $\beta_p < 1$. At the marked points along these boundaries, the maximum growth rates, $\Gamma_{max}$, were determined.}
\label{fig8}
\end{figure}
In picture \ref{fig8} the points of the experimental measurements taken from figure~\ref{fig7} are presented. We require that along these curves there must be a maximum of instability ($\Gamma = \Gamma_{\max}$) of the fundamental mode, $n=0$, for which $p = p_{\max} = 0.4$ and $M = 6$, $\Delta = 0.5$. From expression $p$ in (\ref{dis}), taking into account that $k_z = k \cos(\theta) = k \ell^{1/2}$, we obtain
\begin{equation}
 \frac{k_z}{\sigma}=\frac{p_{max}}{2\Delta M}\sqrt{\beta+2\alpha}, 
 \ \  \beta=\frac{2}{\beta_{p}}. 
\end{equation}
In figure~\ref{fig9} the dependence of $k_z/\sigma$ on $\beta_p$ along the boundaries is presented. As can be seen, both branches in figure~\ref{fig8} (boundaries $\alpha>1$ and $\alpha<1$) practically coincide. This situation suggests that the limits of the anisotropy region at $\beta_{\parallel p} < 1$ are linked to the instability that emerges in the supersonic flow of the anisotropic plasma within the solar wind, driven by the resonance between eigen oscillations and the shearing flow.
\begin{figure}
  \centering
  \includegraphics[width=0.5\textwidth]{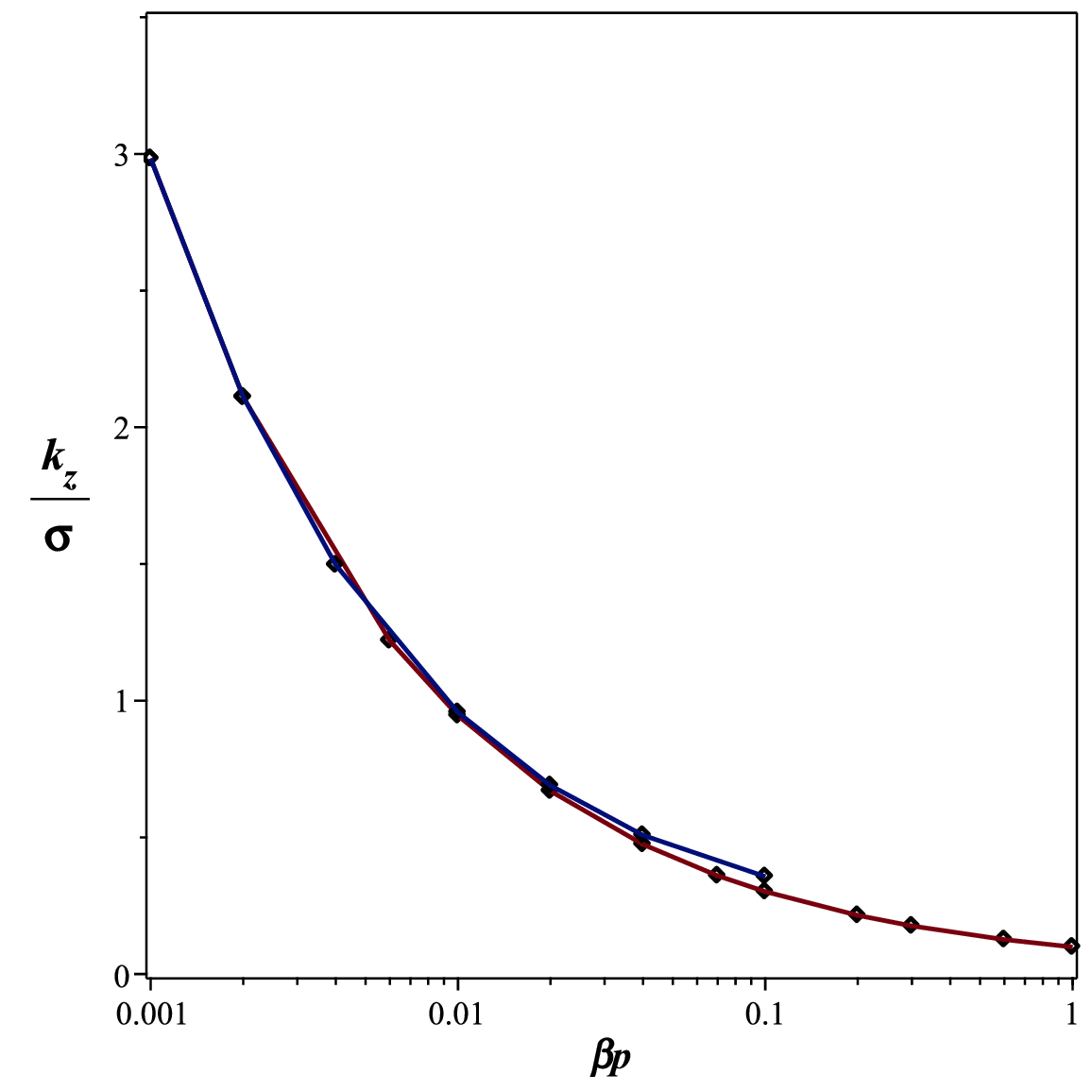}
  \caption{All potential values of $ k_z / \sigma$, representing the ratio of the characteristic scale of flow velocity shear to the wavelength along the flow direction, are determined at the anisotropy boundaries shown in the previous figure \ref{fig8} for the case of maximum growth rate of the fundamental mode with $ n = 0 $. The two plotted curves, corresponding to $ \alpha < 1 $ (red curve) and $ \alpha > 1$ (blue curve), exhibit almost complete overlap. This overlap suggests that, for a given $ \beta_p$, the presence of two distinct boundaries at $ \alpha < 1 $ and $ \alpha > 1 $ is associated with the instability of the same mode of shear flow. }
\label{fig9}
\end{figure}

\begin{figure}
  \centering
  \includegraphics[width=0.5\textwidth]{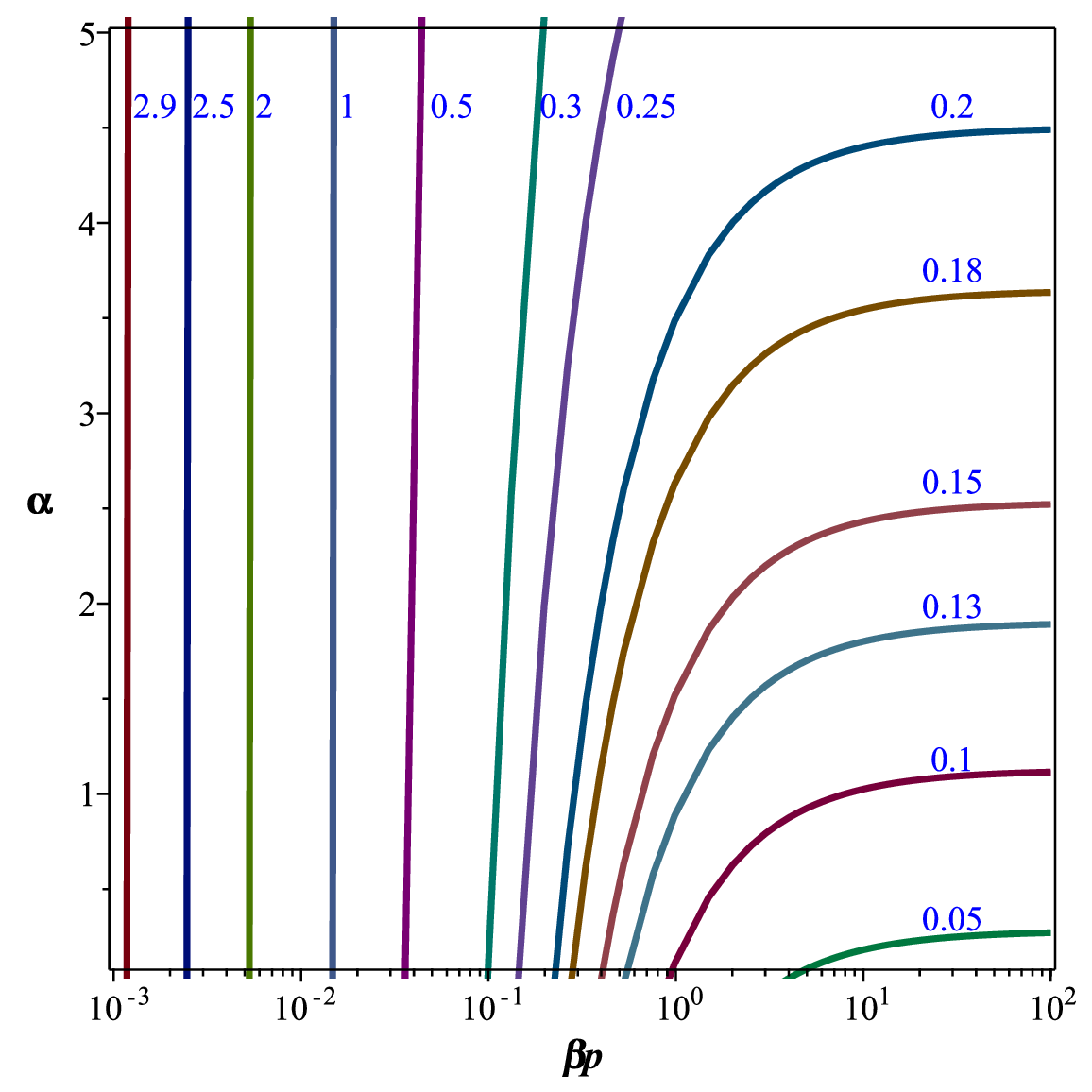}
  \caption{Possible values of the ratio $k_z/\sigma$ (numbers indicated on the curves) within the observed anisotropy range ($\beta_p < 100$, $\alpha < 5$), for which the growth rate of the fundamental mode ($n=0$) reaches its maximum value. Outside the interval $0.01 < k_z/\sigma < 3$, the shear-flow instability does not occur. }
\label{fig10}
\end{figure}
In figure~\ref{fig10} all possible values of $k_z/\sigma$ are shown in the plane of all the observed values $\alpha < 5$ and $\beta_p < 100$. We see that the maximum instability is possible within the limits $ 10^{-2} < {k_z}/{\sigma} < 3.$
This means that for $\sigma \to \infty$ (vortex sheet approximation) and for $\sigma \to 0$ (absence of shear in the flow), the instability found in anisotropic plasma disappears.

\section{Conclusions}

The presence of low-frequency, large-scale Alfv\'enic turbulence within the solar wind plasma is well established (e.g., \citealp{Coleman1968,Belcher1971}). These fluctuations are fundamental to the energetics of the heliosphere, providing the requisite momentum and energy flux for particle acceleration and the maintenance of supersonic expansion. Despite their ubiquity, the specific mechanisms governing collisionless dissipation and the subsequent heating of the coronal and solar wind plasmas remain a subject of active investigation. The transition of energy from large-scale turbulence to dissipative scales necessitates a nonlinear cascade, likely mediated by plasma instabilities that facilitate the development of small-scale structures.
Observational evidence increasingly supports the existence of such inhomogeneities; specifically, periodic density structures have been detected that propagate from the solar corona to the inner heliosphere (\citealp{Viall2021}). These compressional oscillations, typically associated with the slow solar wind, exhibit characteristic periods in the range of 4--140 min, with corresponding radial scales between $8\cdot10^{4}$ and $3\cdot10^{6}$ km. Furthermore, these periodic variations in the proton number density have been observed to persist into the distal solar wind beyond 1 AU (\citealp{Birch2021}).

As noted in the preceding introduction, weakly collisional astrophysical plasmas immersed in a strong magnetic field are prone to developing significant pressure anisotropy relative to the local magnetic field vector. Pressure anisotropy and predominantly field-aligned heat fluxes provide a free-energy source for the excitation of various MHD instabilities, including mirror and firehose modes, as well as other low-frequency kinetic analogs (e.g., \citealp{Kuznetsov2009,Dzhalilov2011,Dzhalilov2013,Hunana2017}).

In addition, in the context of the bimodal solar wind, solar rotation facilitates the formation of stream interaction regions. These interfaces are characterized by supersonic shear flows, which, when coupled with anisotropy and heat-flux effects, act as a catalyst for nonlinear energy transfer. Specifically, these instabilities facilitate the cascade from large-scale Alfv\'enic oscillations toward smaller scales where dissipative processes become dominant (e.g., \citealp{Goldreich1995,Howes2008,Alexandrova2013}).The present study provides a fluid-theoretical analysis of these instabilities to characterize their role in the energetic evolution of the plasma.
Shear flows also serve as a source of free energy capable of driving a broad spectrum of plasma instabilities. In the presence of supersonic shear, specific phenomena emerge, including the KHI type and the excitation of negative-energy waves (e.g., \citealp{Bers2016,Yu2020}). Under these conditions, the Doppler-shifted frequency in the plasma rest frame,
$\omega_{D} = \mathrm{Re}(\omega) - k \cdot V_{0}(x)$ may undergo a sign reversal, leading to over-reflection and wave amplification. When $\omega_{D}<0$, waves propagate backward in the flow frame; in such cases, the wave energy density can become negative, allowing the shear to facilitate a non-adiabatic energy transfer from the bulk flow to the fluctuations via wave--flow coupling. A similar physical situation arises at the interfaces between fast and slow solar wind streams. Since the resonant condition occurs when the phase velocity along the flow matches the mean flow velocity ($v_{ph}\approx\overline{V_{0}}$), the Doppler frequency becomes
$\omega_{D} = k_{z}\left(\overline{V_{0}} - V_{0}(x)\right)$.
The critical layer, defined by $\omega_{D}(x_{c})=0$ (as follows from Equation~(\ref{V0}), $x_{c}=0$), corresponds to the location of the development of the maximum instability (see Figure~\ref{fig1}), particularly where the magnetic field aligns with the interface. The superposed-epoch analysis of 27 CIRs at 1 AU confirms that the stream interface typically coincides with the maximum of the plasma vorticity (\citealp{Borovsky2010}).
It is essential to recognize that while shear interfaces potentially align with the Parker spiral in the interplanetary medium where the magnetic field is quasi-parallel to the flow---the Sun's rotation ensures that velocity shear may also develop in the inner heliosphere (\citealp{Pinto2021}).

Furthermore, we have demonstrated that this instability mechanism provides a viable explanation for the observed limits of the proton temperature anisotropy in the low-plasma-beta ($\beta_{p}<1$) regime. This result is of particular significance, as it suggests that in low-$\beta_{p}$ environments, plasma isotropization is governed predominantly by plasma stream effects, particularly in a fluid-dynamic framework. Consequently, these findings offer a potential resolution to a long-standing challenge in solar wind plasma physics regarding the mechanisms of relaxation in weakly collisional regimes.

Presently, our analysis is constrained to a simplified model, where the stationary state of the plasma is assumed to be homogeneous, with the shear flow aligned parallel to a uniform magnetic field. In more physically realistic configurations, the system exhibits increased complexity, characterized by oblique flow orientations relative to a non-uniform magnetic field. Specifically, the angle between the velocity and magnetic field vectors may increase with heliocentric distance, potentially reaching a perpendicular orientation in the distal solar wind. Also, in such collisionless regimes, the characteristic ion kinetic scales undergo significant expansion, eventually becoming comparable to the shear flow scale. Consequently, the standard fluid equations must be modified to account for these finite-scale effects. Our current formulation employs a system of equations derived for a shifted bi-Maxwellian plasma. While this approach yields a uniform stationary equilibrium even when longitudinal heat flux is included (neglecting self-gravity), such homogeneous solutions are found to be non-existent under more generalized, modified steady-state conditions (e.g., \citealp{Cerri2014,Malara2022}).

The investigation of the nonlinear evolution of shear-driven instabilities in anisotropic plasmas within a fluid-theoretical framework constitutes a primary objective for subsequent research. \citet{Henri2013} conducted a comparative analysis of various closures---including MHD, Hall-MHD, multi-fluid, and hybrid/full kinetic models---to evaluate their impact on the Kelvin--Helmholtz instability in both linear and nonlinear regimes. A central aim of that study was to determine the role of microphysical processes in governing the macroscale dynamics of collisionless plasmas.Their findings demonstrate that large-scale kinetic simulations largely validate the fluid-dynamic approximation, confirming the utility of fluid codes for investigating the nonlinear behavior of magnetized flows, even under collisionless conditions. Consistent with the predictions of kinetic theory, it is concluded that fluid-based methodologies remain robust for resolving the large-scale phenomenology of these systems.
\section*{Declaration of Interests}
The authors report no conflicts of interest.

\bibliographystyle{spr-mp-sola}
\bibliography{sola_bibliography_example}
\end{document}